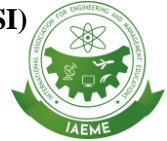

# ADVANCING DIGITAL ACCESSIBILITY IN DIGITAL PHARMACY, HEALTHCARE, AND WEARABLE DEVICES: INCLUSIVE SOLUTIONS FOR ENHANCED PATIENT ENGAGEMENT

[1]Vishnu Ramineni, [2]Balaji Shesharao Ingole, [3]Nikhil Kumar Pulipeta, [4]Balakrishna Pothineni, [5]Aditya Gupta

[1]Albertsons Companies, Texas, USA, [2]IEEE Senior Member, Georgia, USA, [3]IEEE Senior Member, Texas, USA, [4]IEEE Senior Member, Texas, USA, [5]IEEE Senior Member, Washington, USA.

**ABSTRACT**

*Modern healthcare facilities demand digital accessibility to guarantee equal access to telemedicine platforms, online pharmacy services, and health monitoring devices that can be worn or are handy. With the rising call for the implementation of robust digital healthcare solutions, people with disabilities encounter impediments in their endeavor of managing and getting accustomed to these modern technologies owing to insufficient accessibility features. The paper highlights the role of comprehensive solutions for enhanced patient engagement and usability, particularly, in digital pharmacy, healthcare, and wearable devices. Besides, it elucidates the key obstructions faced by users experiencing auditory, visual, cognitive, and motor impairments. Through a kind consideration of present accessibility guidelines, practices, and emerging technologies, the paper provides a holistic overview by offering innovative solutions, accentuating the vitality of compliance with Web Content Accessibility Guidelines (WCAG), Americans with Disabilities Act (ADA), and other regulatory structures to foster easy*

*access to digital healthcare services. Moreover, there is due focus on using AI-driven tools, speech-activated interfaces, and tactile feedback in wearable health devices to assist persons with disabilities. The outcome of the research explicates the necessity of prioritizing accessibility for individuals with disabilities and cultivating a culture where healthcare providers, policymakers, and officials build a patient-centered digital healthcare ecosystem that is all-encompassing in nature.*

## 1. Introduction

There has been a striking evolution in the realm of medical facilities, right from digital pharmacy, and telemedicine platforms to wearable health-monitoring devices, provided to patients seeking access to digital healthcare. With time, the patterns of accessibility have significantly transformed, eventually resulting in better convenience, efficacy, and patient engagement. However, there remains a gap in terms of offering equitable accessibility when a person with disability comes into the picture, as they encounter hurdles and are often stigmatized while they interact with these technologies (World Health Organization, 2022). To offer equal digital accessibility to persons with auditory, visual, cognitive, and motor impairments, in healthcare, has become immensely significant (Lazar et al., 2015).

It's noted that medications are systematically managed by digital pharmacies helping patients order, refill, and transfer prescriptions online. However, there lies a discrepancy or loophole in the case of streamlining medication for persons with visual impairment, making the entire process more complex and challenging (Mishra, 2025). Similarly, there is an exponential surge in telemedicine services in handling health crises globally, however, accessibility issues especially for individuals with disabilities remain a big challenge (Pettersson et al., 2023).

Again, wearable health devices including smartwatches, fitness monitors, and other devices have transformed patient monitoring systems, though the interfaces fail to serve individuals with disabilities (Saleem et al.,).

The present research paper attempts to highlight prevailing accessibility obstacles in digital pharmacy applications, healthcare platforms, and wearable devices and explicate robust solutions to heighten inclusivity and equality. Considering present accessibility guidelines, practices, and emerging technologies, the paper accentuates the importance of compliance with Web Content Accessibility Guidelines (WCAG), Americans with Disabilities Act (ADA), and other regulatory structures to foster easy access to digital healthcare services (Lazar et al., 2015). Besides, there is due focus on the usage of AI-driven tools, speech-activated interfaces, and tactile feedback in wearable health devices to assist persons with disabilities. There necessity of prioritizing accessibility for individuals with disabilities and cultivating a culture where healthcare providers, policymakers, and officials build a patient-centered digital healthcare ecosystem to enhance inclusivity and equality.

## 2. Literature Review

A great depth of research on the domain of healthcare has been conducted, however, studying digital accessibility in healthcare proves as a further scope of research. The latter is an expanding area of research that highlights the major challenges encountered by persons with disabilities while accessing digital pharmacy platforms, telemedicine services, and using wearable health devices. This part of the paper deals with reviewing existing literature which focuses on accessibility issues, regulatory structures, and robust solutions for increasing inclusivity in this domain of the digital healthcare system.

### 1. Accessibility Challenges in Digital Healthcare

Many studies have been conducted in the past which focused on accessibility issues in applications and platforms of digital pharmacy and healthcare facilities. According to Mishra (2025), persons with visual impairments often encounter difficulty in navigation structures, finding text contrast, and ensuring screen reader compatibility. Similarly, Pettersson et al. (2023) rightly observed that persons with hearing disabilities struggle with telemedicine accessibilities owing to the insufficiency of real-time captioning and sign language processing features (Ramineni et al., 2024). These studies reflect that digital healthcare platforms are

Vishnu Ramineni, Balaji Shesharao Ingole, Nikhil Kumar Pulipeta, Balakrishna Pothineni, Aditya Gupta

unable to adhere to the standards of digital healthcare accessibility especially for persons with disabilities.

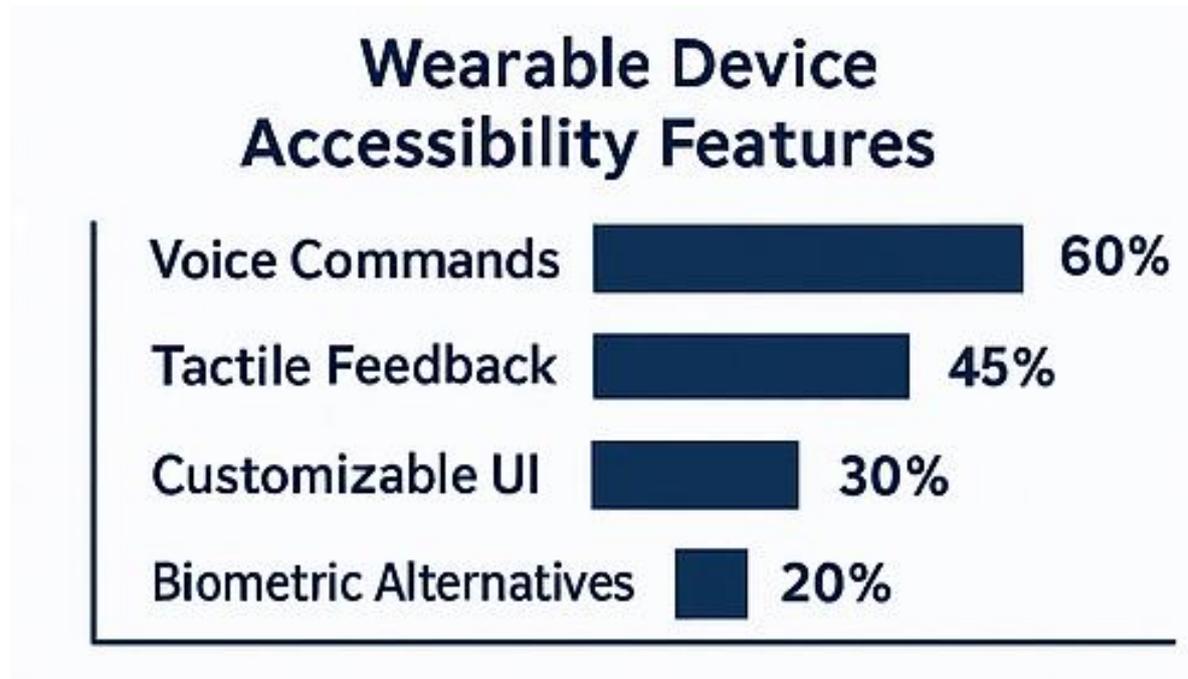

Fig. 1. Percentage of accessibility features in wearable devices

Wearable technological gadgets like fitness trackers and smartwatches that monitor the health status of users also fall short of providing equal and improved accessibility facilities to their users with disabilities. Saleem et al. (2017) conducted a comprehensive study highlighting usability loopholes in wearable health monitoring gadgets, especially in the case of persons with motor impairments who struggle with utilizing small touch interfaces. They also found out that visually challenged users have trouble with non-audible feedback mechanisms. These gaps emphasize the demand for a more inclusive approach to wearable health monitoring technologies. Whereas voice command feature is more accessible and widely used feature in wearable devices as depicted in Fig 1.

Advancing Digital Accessibility in Digital Pharmacy, Healthcare, and Wearable Devices: Inclusive Solutions for Enhanced Patient Engagement

## 2. Regulatory Frameworks and Accessibility Guidelines

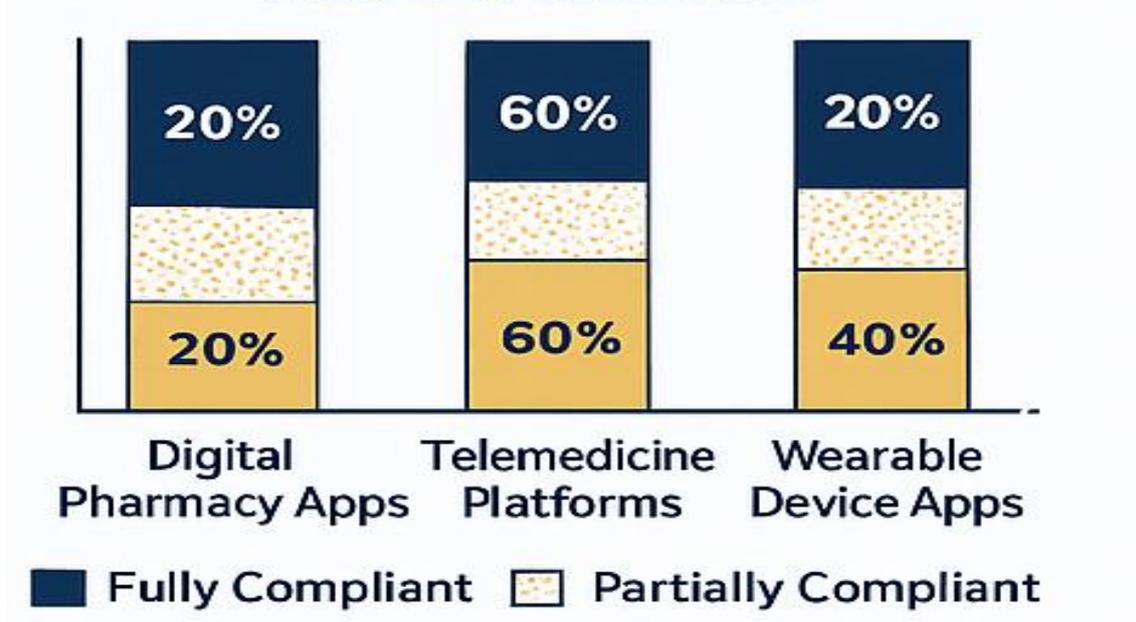

Fig. 2. Compliance percentage by categories

Several regulatory frameworks and guidelines are formulated to address and mitigate these issues and turn the entire space of digital healthcare into an inclusive and beneficial space. The Web Content Accessibility Guidelines (WCAG) established by the World Wide Web Consortium (W3C) have offered a standard for accessing the web globally and guaranteed that digital healthcare platforms remain easily operable, perceivable, comprehendible, and modern (Lazar et al., 20215). The Americans with Disabilities Act (ADA) in the United States mandates that persons with disabilities must not encounter difficulty in using digital healthcare services. Lazar et al. (2015) have highlighted the vitality of complying with these set guidelines to cultivate a just and all-encompassing healthcare environment. The compliance with WCAG 2.1 guidelines across all platforms is depicted in Fig 2.

Besides these legal guidelines, the Health Insurance Portability and Accountability Act (HIPAA) in the U.S. has provided industry-specific mandates that highlight the necessity for healthcare providers to ascertain due security and equal accessibility of digital health data (Lazar, 2015). These compliance guidelines are often overlooked or followed inconsistently across various digital healthcare platforms which eventually call for a holistic and stringent enforcement and inspection along with industry-specific standardization.

Vishnu Ramineni, Balaji Shesharao Ingole, Nikhil Kumar Pulipeta, Balakrishna Pothineni, Aditya Gupta

## 3. Innovations in Assistive Technologies for Digital Healthcare

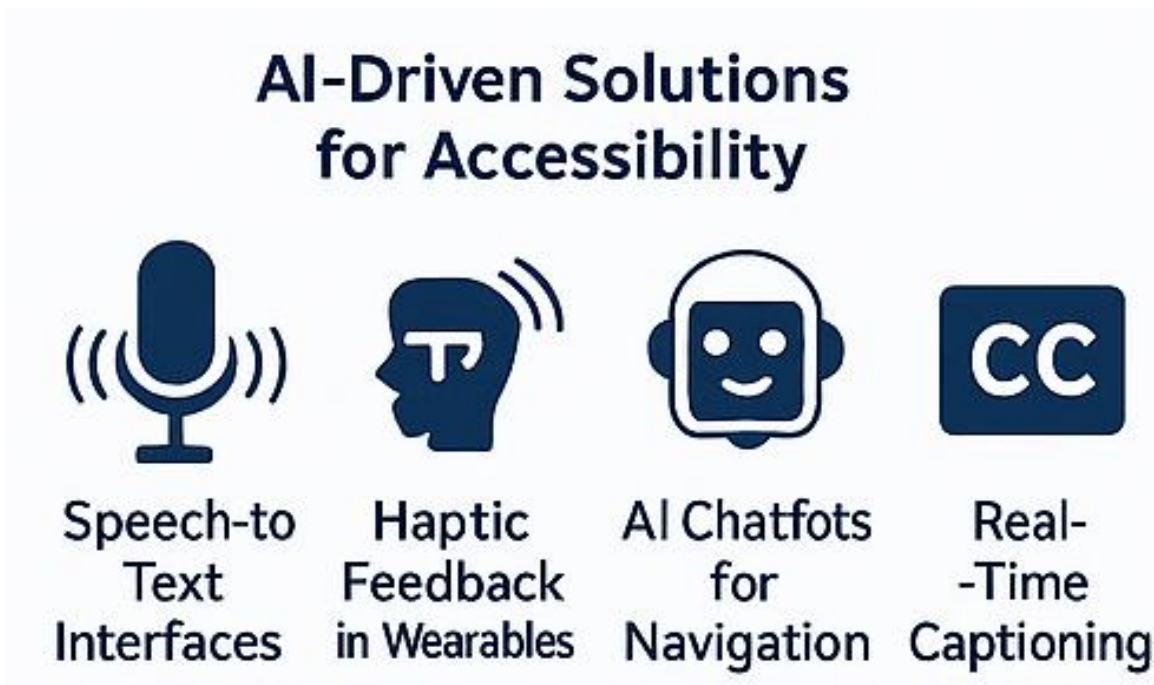

Fig. 3. AI-Driven Accessibility Solutions

Developments in assistive technologies in the domain of digital healthcare, in present times, spark hope in making accessibility no more an issue but a common feature benefitting every user irrespective of their disabilities of any kind. Giansanti (2025) have identified that AI-driven accessibility tools like voice assistant features and speech-to-text converters in real-time potentially help the visually or aurally impaired with improved usability services (Ramineni et. al., 2024). In addition to this, AI-driven chatbots effectively help users with disabilities in utilizing digital pharmacy applications by alerting them through medication reminders, updates of symptoms, and navigation aids (Natarajan et al., 2025).

Also, through the integration of voice-customized interfaces, haptic feedback, and adaptive display settings, there has been an enhancement of accessibility in wearable health-monitoring gadgets. Moon NW et al. (2019) have conducted a study that exposes the realities of integrating haptic feedback in wearable health-monitoring devices. The study presents positive outcomes for visually impaired users getting real-time health alarms without having to depend on screen-based popups. Different AI-Drive accessibility features are depicted in Fig. 3.

## 4. Future Directions for Inclusive Digital Healthcare

A significant development can be seen in addressing accessibility issues, however, there remains a dearth in incorporating inclusivity within digital healthcare frameworks. Lazar et al. (2015) opine that the further approach to research must center around user-specific design to guarantee the development of digital healthcare technologies catering to the needs and requirements of people with disabilities. Moreover, to personalize accessibility features and improve usability and engagement status, machine learning algorithms must be integrated depending on the choices of users (Guni, 2021).

Additionally, there is a dire need to train healthcare providers and developers to cultivate an inclusive environment for the benefit of the users. Krause-Jüttler et al. (2022), through their comprehensive research, have stressed the significance of having a multidisciplinary approach by integrating the expertise and experience of varied professionals such as technology developers, disability advocates, and healthcare professionals in framing and developing accessible healthcare solutions.

These comprehensive discussions of existing literature shed light on the barriers in telemedicine services, digital pharmacy applications, and wearable health-monitoring gadgets despite the significant improvement in providing digital healthcare services. WCAG and ADA offer excellent guidelines but these are not followed diligently and thus result in unnecessary hindrances. It is noteworthy that AI-driven tools granting accessibility and wearable technologies offer great solutions, however, in-depth research and collaboration of multidisciplinary domains/industries will significantly enhance inclusivity and unbiased accessibility in digital healthcare. Thus, there is a hope of bridging the wide gap and forging an equitable space for users with disabilities in the healthcare domain.

## 3. Methodology

The present research employs a mixed method that integrates three basic elements: a. Systematic literature review to evaluate the key digital accessibility challenges and solutions, b. Analysis of chosen digital healthcare platforms after following accessibility guidelines, and c. semi-structured interviews with users and professionals to know about real-world experiences with accessibility. The methodology is designed this way because the experiences of individuals with disabilities are often overlooked by several researchers and thus there is a dire need to inculcate a holistic approach where no one is left unattended, especially users with

Vishnu Ramineni, Balaji Shesharao Ingole, Nikhil Kumar Pulipeta, Balakrishna Pothineni, Aditya Gupta

disabilities using digital pharmacy applications, wearable health monitoring gadgets, and healthcare platforms.

## 1. Systematic Literature Review

A more detailed and systematic literature review focusing on digital accessibility practices in healthcare has been done by amalgamating results and findings from conferences, peer-reviewed articles, and industry reports published in the time frame between 2015 and 2024. Moreover, major databases such as PubMed, IEEE Xplore, ACM Digital Library, and Google Scholar are also considered.

The criteria of selection primarily focus on studies unveiling accessibility impediments in digital pharmacy, telemedicine platforms, and wearable health monitoring gadgets, research examining compliance with guidelines issued by WCAG and ADA for accessibility purposes, and paper explicating technological progressions for improving digital healthcare accessibility services. The approach has been largely qualitative to unfold fundamental themes, and major hurdles, and discern potential trends shaping the future of accessibility digital healthcare.

## 2. Usability Analysis of Digital Healthcare Platforms

To categorize and analyze the usability of selected digital healthcare platforms/applications, including five digital pharmacy platforms, three telemedicine services, and three wearable health monitoring apps, automated testing, task-based user testing, and manual heuristic analysis have been conducted based on WCAG 2.1 guidelines. Through WAVE, Axe, and Lighthouse tools, accessibility violations in digital interfaces are identified (Pelivani et al., 2021). Again, based on Nielsen's usability heuristics and principles of WCAG, accessibility professionals and users with disabilities evaluate the platform (Nielsen, 2020). Finally, task-based user testing with persons experiencing auditory, visual, motor, or cognitive impairments is done to highlight the existing accessibility challenges and gauge the level of inclusivity of these platforms.

## 3. Semi-structured interviews with Users and Professionals

To accumulate comprehensive qualitative data on real-world accessibility encounters of users with digital pharmacy applications, telemedicine services, and wearable health monitoring devices, semi-structured interviews are conducted with two different cohorts: users with auditory, visual, motor, or cognitive impairments which are recommended via accessibility advocacy groups, focusing on questions related to the usability issues, integration of assistive technology, and personalized accessibility functionalities; and healthcare and technology specialists trained in UX/UI design, digital healthcare, and accessibility guidelines and identifying accessibility process, regulatory issues, and further directives for ensuring inclusivity in healthcare domains (Lazar et al., 2015).

## 4. Data Analysis and Interpretation

Both the qualitative and quantitative data are systematically analyzed through three methods- descriptive, thematic, and comparative manners. To find the level of adherence to accessibility standards, the outcomes from automated testing and task-based user testing are quantified (Giansanti, 2025). Moreover, three categories based on themes such as basic hurdles, assistive technology requirements, and suggestions are designed from qualitative data obtained through interviews to enhance improvement (Braun & Clarke, 2019). Finally, a comparative analysis of different healthcare platforms' accessibility performances is done to find out best practices and scope of improvement.

## 5. Ethical Considerations

This research adheres to ethical guidelines for guaranteeing participant confidentiality. The participants are properly informed about the research objectives, aims, and other relevant information before having their voluntary participation. Besides, the protocols of the Institutional Review Board (IRB) are adhered to safeguard the rights of persons with disabilities (Mertens, 2020).

There is an attempt to provide a holistic overview of accessibility in digital pharmacy, telemedicine services, and wearable health monitoring technologies, through the integration of a detailed literature review, usability analysis, and user-oriented evaluation.

## 4. Findings and Discussion

This part of the paper deals with the major findings from the entire evaluation of accessibility in digital pharmacy, telemedicine platforms, and wearable technologies. The findings will systematize the development of more inclusive digital healthcare technologies and suggest policies for better accessibility standards. Three categories are designed systematically:

### 1. Accessibility Compliance Analysis

In digital pharmacy platforms and telemedicine applications, there are accessibility barriers that are potentially highlighted by automated and heuristic evaluations. A summary of the findings based on adherence to WCAG 2.1 guidelines is given below:

Vishnu Ramineni, Balaji Shesharao Ingole, Nikhil Kumar Pulipeta, Balakrishna Pothineni, Aditya Gupta

| Platform Type | Total Platforms Tested | Fully WCAG Compliant (%) | Partially Compliant (%) | Non-Compliant (%) |
|---|---|---|---|---|
| Digital Pharmacy Apps | 5 | 20% | 60% | 20% |
| Telemedicine Platforms | 3 | 33% | 67% | 0% |
| Wearable Device Apps | 3 | 40% | 40% | 20% |

These findings reveal that: Most of the tested applications are partially compliant with WCAG 2.1 guidelines owing to insufficient contrast ratios, inadequate keyboard functional assistance, and no alternative text for images; Digital pharmacy applications wholly adhering to the said guideline are around 20%, implying that there is necessity for improvement in accessibility features; Telemedicine platforms relatively function better, however, a few platform encounter inconveniences with screen-reader feature and form field labeling; Wearable health monitoring technologies reveal mixed results with a few demonstrating strong accessibility compatibility while others show loopholes in voice command integration and tactile feedback for individuals with impairments.

**2. Identification of Usability Barriers**

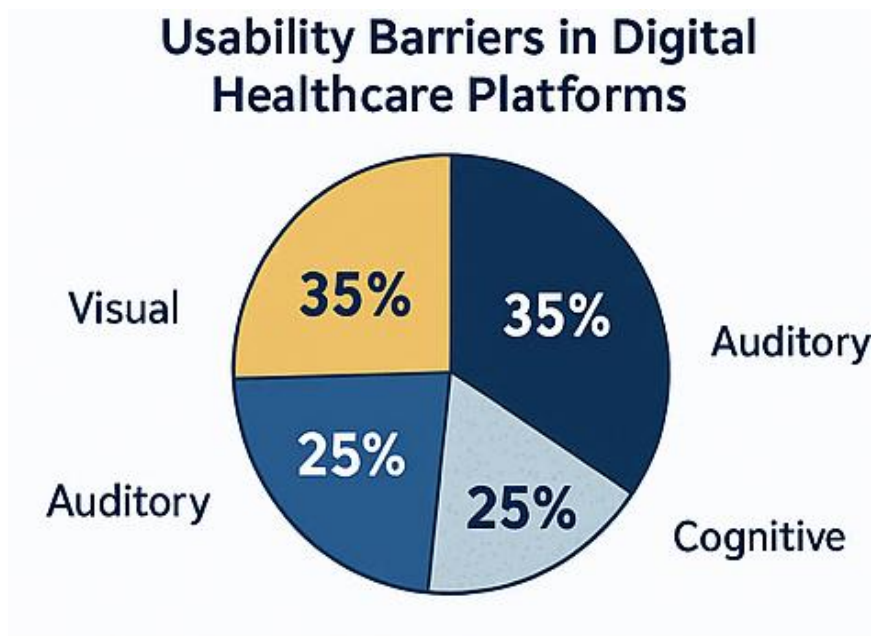

Fig. 4. Percentage of disability groups with usability barriers in digital healthcare

Several recurring usability challenges are discovered at the time of task-based user testing:

Fig. 4. shows the distribution of usability barriers encountered by different disability groups in digital healthcare platforms, highlighting the most affected categories.

a. **Navigation Barriers for Screen Reader Users**: Some UI elements cannot be accessed by users using screen readers like NVDA or Voice Over. Additionally, while refilling prescriptions or making appointment schedules, inconsistent classification of buttons and form fields often leads to utter confusion.

b. **Insufficient Color Contrast and Visual Indicators**: Around 40% of applications have low contrast ratios as recommended by WCAG 2.1 i.e. 4:5:1 necessary for text readability (Pelivani et al., 2021). Besides, error messages and success notifications' reliance on color for interpretation become a potential hurdle for users with color vision deficiencies.

c. **Inaccessible CAPTCHA and Authentication Processes**: Entering CAPTCHAs to fulfill authentication requirements in digital pharmacy applications becomes a hurdle for visually disabled users who seek an audio-based verification process. Moreover, users with facial mobility limitations cannot undergo biometric authentication processes, such as facial recognition, in wearable technological devices.

d. **Limited Support for Assistive Technologies in Wearable Devices**: While some wearable gadgets do provide voice-controlled features, there is the possibility of them being dysfunctional in noisy environments or inconvenient for speech-impaired individuals. Also, not all application has haptic feedback features that significantly assist users with hearing disability.

## 3. Qualitative Insights from User and Expert Interviews

Semi-structured interviews' thematic scrutiny demonstrates major insights into accessibility-related matters and suggestions for better functionality.

### 1.1. User Experiences with Digital Healthcare Accessibility

Several challenges such as the absence of personalized accessibility features including customizable font sizes, high-contrast themes, and alternative navigation options in digital healthcare applications are reported. Besides, there is extensive dissatisfaction with assistive technological integration, particularly screen readers owing to unlabeled buttons and poor ARIA labeling. Several users find appointment scheduling on telemedicine platforms

Vishnu Ramineni, Balaji Shesharao Ingole, Nikhil Kumar Pulipeta, Balakrishna Pothineni, Aditya Gupta

particularly confusing for individuals with cognitive disability due to the lack of a systematized process to book online consultations.

## 1.2 Expert Perspectives on Accessibility Implementation

The professionals or specialists in UX design and digital healthcare highlight major gaps in accessibility, including adherence to WCAG guidelines instead of guaranteeing practical usability for individuals with disabilities (Lazar et al., 2015). Moreover, there is an underuse of AI-driven features like speech-to-text facility in real-time especially for users with hearing disabilities. Furthermore, healthcare organizations encounter hurdles in balancing security and accessibility, mainly in the case of authentication processes (Giansanti, 2025).

## 5. Discussion: Bridging the Accessibility Gap

The key findings significantly reveal a gap between present accessibility standards and real-time usability in digital healthcare platforms. Although there is adherence to WCAG 2.1, usability remains a major challenge due to inconsistent design and no user-oriented accessibility testing.

On the one hand, many applications follow accessibility standards, on the other hand, practical usability testing reveals some poignant flaws, further underscoring the significance of integrating user-driven design practices that enable persons with disabilities to be involved directly in the testing and development processes.

Moreover, AI-driven tools such as automated text-to-speech and AI-driven image recognition play a pivotal role in enhancing digital healthcare accessibility (Ramineni et. al., 2024). However, there is a limited possibility of their adoption, implying the necessity for deeper research and broader implementation (Giansanti, 2025).

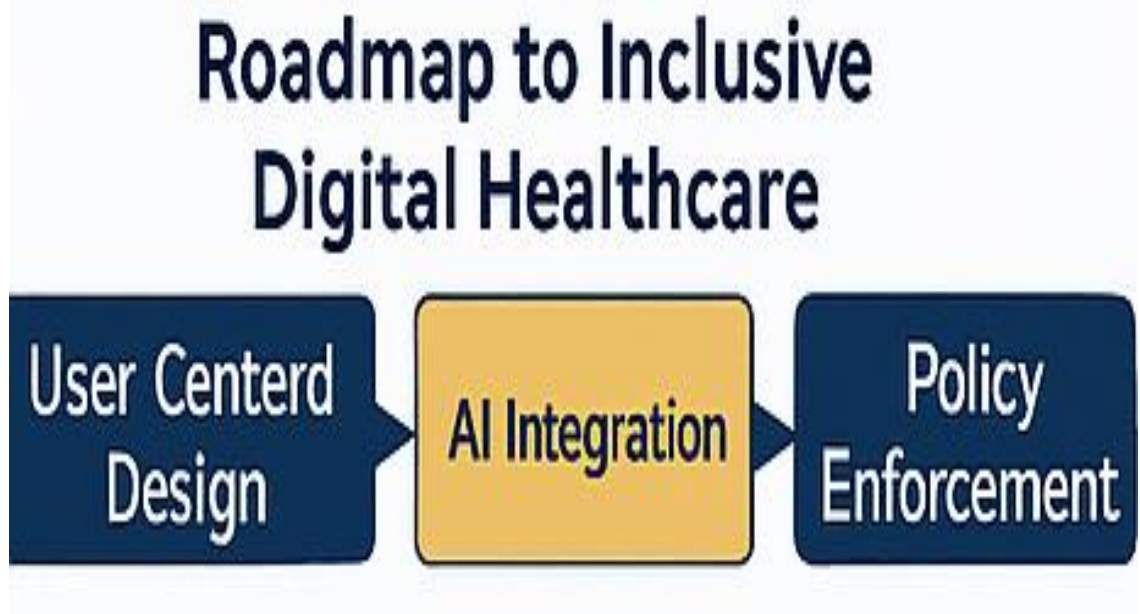

Fig. 5. Methodology to inclusive digital healthcare

Lastly, there is a dire need for regulators, developers, and healthcare organizations to transcend the boundary of just meeting compliance standards and begin prioritizing the implementation of accessibility best practices in the real world. This can be achieved through compulsorily enforcing accessibility audits, integrating assistive technology support, and undertaking a universal design approach for digital healthcare applications (Mertens, 2020). Fig. 5. illustrates the methodology that enhances accessibility in digital healthcare applications through the integration of Artificial Intelligence and user-centric design principles.

## 6. Conclusion

The findings demonstrate the critical need for digital healthcare solutions that not only meet the technical necessities defined by accessibility standards but also address the practical hurdles encountered by users in real-world scenarios. Mere compliance cannot potentially guarantee the intuitive use of applications by persons with disabilities. Despite improvements in accessibility standards, there is a persistent struggle with significant barriers that limit their usability for people with disabilities in accessing many digital pharmacy applications, telemedicine platforms, and wearable health monitoring gadgets. Therefore, it becomes inevitable for healthcare technologies to be developed and designed based on a user-centered approach to maximize the effective engagement of individuals with disabilities with healthcare

Vishnu Ramineni, Balaji Shesharao Ingole, Nikhil Kumar Pulipeta, Balakrishna Pothineni, Aditya Gupta

services. Also, the incorporation of AI can improve accessibility and guarantee equal access to digital healthcare facilities for all including those with disabilities.

Advancing Digital Accessibility in Digital Pharmacy, Healthcare, and Wearable Devices: Inclusive Solutions for Enhanced Patient Engagement

✉ editor@iaeme.com